\documentclass[11pt,preprintnumbers,aps,amssymb,nofootinbib,amsmath,superscriptaddress]
{revtex4}
\usepackage{epsfig,epsf}
\usepackage{bm} 
\usepackage{color} 
\newcommand{\beq}{\begin{equation}}
\newcommand{\beql}[1]{\begin{equation}\label{#1}}
\newcommand{\eeq}{\end{equation}}
\def\bal#1\gal{\begin{align}#1\end{align}}
\newcommand{\ball}[1]{\bal\label{#1}}
\newcommand{\eq}[1]{(\ref{#1})}
\newcommand{\fig}[1]{Fig.~\ref{#1}}
\renewcommand{\sec}[1]{Sec.~\ref{#1}}
\newcounter{topiccounter}
\renewcommand{\b}[1]{{\bm #1}} 
\newcommand{\unit}[1]{\hat {{\bm #1}}} 

\begin{document}

\title{Initial value problem for magnetic field in heavy ion collisions}

\author{Kirill Tuchin}

\affiliation{
Department of Physics and Astronomy, Iowa State University, Ames, Iowa, 50011, USA}

\date{\today}


\begin{abstract}

When quark-gluon plasma emerges in the wake of a heavy-ion collision, magnetic field created by the valence charges has already permitted the entire interaction region. Evolution of this ``initial" field in plasma is governed by Maxwell equations in electrically conducting medium. As the plasma expands external valence charges induce magnetic field that also contributes to the total magnetic field in plasma. We solve the initial value problem describing these processes and argue that the initial magnetic field often dominates over the one induced by the valence charges. In particular, it grows approximately proportional to the collision energy, unlike the induced component, which is energy-independent. As a result, magnetic field has a  significant  phenomenological  influence on quark-gluon plasma at the LHC energies over its entire lifetime.

\end{abstract}

\maketitle

\section{Introduction}\label{sec:a}

In this paper we revisit the problem of magnetic field  created by  electrical currents of colliding relativistic heavy ions \cite{Tuchin:2010vs,Voronyuk:2011jd,Skokov:2009qp,Deng:2012pc,Bzdak:2011yy,McLerran:2013hla,Tuchin:2013apa,Zakharov:2014dia}. Since these currents experience very little  deflection in the course of collision \cite{Itakura:2003jp,Kharzeev:1996sq}, the corresponding magnetic field depends  on energy and geometry of the collision, and implicitly on the strong interaction dynamics through the   electrical conductivity of the quark-gluon plasma (QGP) \cite{Tuchin:2010vs,Tuchin:2013apa}. 
Another important aspect, which is the main focus of this study, is transition dynamics from magnetic field in vacuum to the one in medium. To begin with,  assume that QGP forms instantly at time $t=t_0$, where $t$ is counted from the collision time in the laboratory frame. This time emerges in phenomenological models of QGP  that  favor rather  small values as compared to the perturbation theory expectations, see e.g.\ \cite{Kolb:2003dz}. The earliest  possible value  of $t_0$ is determined by the saturation momentum $Q_s$ as $1/Q_s$ and represents the time it  takes to release most particles from the ion's wave functions. At RHIC  $1/Q_s\sim 0.2$~fm.  At $t<t_0$ we are dealing with electromagnetic field created by the valence charges in vacuum. Its magnetic component is given by the well-known formula \eq{a23}. At time $t=t_0$, when the QGP emerges, magnetic field permits the entire plasma. Starting at $t=t_0$ and on behavior of magnetic field is governed by the Maxwell equations in plasma. These equations describe evolution of magnetic field  in electrically conducting QGP starting from its initial value at $t=t_0$. This component of the total magnetic field is referred to below as the ``initial" magnetic field $\b B_\text{init}$. Another contribution to magnetic field is induced by  valence charges moving outside of QGP and is referred to below  as the ``valence" contribution $\b B_\text{val}$.\footnote{To avoid confusion we emphasize that both components are ultimately related to electrical charges of heavy-ions. The distinction only concerns our treatment of magnetic field at $t>t_0$ as will be explained in detail in the forthcoming sections.} In previous publications the role of the initial field has not been properly recognized. In this paper we fill this void and moreover, argue that in most cases the main contribution stems  from the initial field. 

The paper is organized as follows. In \sec{sec:x}--\sec{sec:b} we deal with magnetic field produced by a single point charge. In \sec{sec:x} we consider magnetic field in vacuum, while in later sections -- in electrically conducting QGP.  The main result is given by equations \eq{b27}, \eq{b25} which represent contributions of valence charges and the initial field respectively. A more realistic geometry is considered in \sec{sec:c} where we discuss the case of two electric charges colliding at a given impact parameter $b$. We also discuss there the effect of time dependent electrical conductivity on the magnetic field evolution. We discuss the results and summarize in \sec{sec:f}.

\section{Magnetic field in vacuum}\label{sec:x}

In a relativistic heavy-ion collision, electromagnetic field is created by $Z$ electric charges of one ion and $Z$ electric charges of another ion moving in the opposite directions along, say, $z$ axis such that ion centers are at a distance $b$ away. Due to the superposition principle, the total classical field is a sum of fields of all charges. Thus, in order to find the total field it is sufficient to solve for a single electric charge $e$. In this section we briefly review a textbook case of electromagnetic field created in vacuum  by a uniformly moving point charge $e$. Our intent here is to introduce notations, definitions etc.

Before the QGP formation, viz.\ at $t\le t_0$, the vector potential $\b A_1(\b r, t)$  of a point charge $e$ moving along the trajectory $z=vt$ satisfies the following equation 
\ball{a11}
\nabla^2\b A_1(\b r, t)= \partial_t^2 \b A_1(\b r, t)-\b j(\b r, t)\,,
\gal
where the electromagnetic current density due to a valence charge $e$ is 
\ball{a13}
\b j = ev\unit z \delta(z-vt)\delta(\b b)\,.
\gal
The momentum space representation is defined as follows
\ball{a15}
\b j(\b r, t)= \int \frac{d^3k}{(2\pi)^3}e^{ i\b k\cdot \b r}\b j_{\b k\omega}= \int \frac{d^2k_\bot dk_z}{(2\pi)^3}e^{ i\b k_\bot\cdot \b b+ik_z z}\b j_{\b k\omega}\,.
\gal
With this normalization the Fourier component of the current reads
\ball{a17}
\b j_{\b k}= e v \unit z e^{-ik_z v t}\,.
\gal
It follows from \eq{a11} that the vector potential generated by the current \eq{a17} is 
\ball{a19}
\b A_{1\b k}= \frac{2\pi e v\unit z}{k^2-k_z^2v^2}=\frac{2\pi e v\unit z}{k_z^2/\gamma^2+k_\bot^2}\,.
\gal
In the configuration space we obtain
\ball{a21}
\b A_1(\b r, t) = \frac{\gamma e v \unit z}{4\pi}\frac{1}{\sqrt{b^2+\gamma^2(vt-z)^2}}\,,
\gal
where $\gamma= (1-v^2)^{-1/2}$. The corresponding magnetic field 
\ball{a23}
\b B_1= -\partial_b  A_1\unit \phi= \frac{\gamma e v \unit \phi}{4\pi}\frac{b}{(b^2+\gamma^2(vt-z)^2)^{3/2}}\,.
\gal
This solution is valid until $t=t_0$ at which time existence of  electrically conducting medium must be taken into account. 

\section{Exact solution for constant electrical conductivity}\label{sec:d}

Maxwell equations can be solved exactly for $t\ge t_0$ in the case of constant electrical conductivity $\sigma$.  The vector potential $\b A_2$ satisfies the following equation
\ball{d5}
\nabla^2 \b A_2(\b r, t)= \partial_t^2\b A_2(\b r, t) +\sigma \partial_t \b A_2(\b r, t)-\b j(\b r, t)\,
\gal
with the initial conditions 
\bal
\b A_2(\b r, t_0)&= \b A_1(\b r, t_0)\equiv \unit z \Phi(\b r,t_0)\,,\label{d7}\\
\partial_t{\b A}_2(\b r, t_0)&= \partial_t{\b A}_1(\b r, t_0)\equiv \unit z \Psi(\b r,t_0)\,.\label{d8}
\gal
We assumed that permittivity and permeability of QGP is trivial. One can take a more accurate account of medium properties, which would yield  more elaborate initial conditions. However, they are not expected to significantly change the final result.  

In momentum space Eq.~\eq{d5} and the corresponding initial conditions \eq{d7},\eq{d8} read
\bal
-k^2\b A_{2\b k}(t)&= \partial^2_t \b A_{2\b k}(t)+\sigma \partial_t \b A_{2\b k}(t)-ev\unit z e^{-ik_z vt}\,,\label{d10}\\
\b A_{2\b k}(t_0)&= \unit z\Phi_{\b k}(t_0)= \frac{ev\unit z}{k_z^2/\gamma^2+k_\bot^2}e^{-ik_z v t_0}\,,\label{d12}\\
\partial_t\b A_{2\b k}(t_0)&=\unit z \Psi_{\b k}(t_0)= -ik_z v\frac{ev\unit z}{k_z^2/\gamma^2+k_\bot^2}e^{-ik_z v t_0}\,.\label{d13}
\gal
To solve \eq{d10}, we first consider the corresponding homogeneous equation 
\bal
&-k^2a_{\b k}(t)= \partial^2_t a_{\b k}(t)+\sigma \partial_t a_{\b k}(t)\,,\label{d15}
\gal
Seeking its solution  in the form  $a_{\b k}\propto  e^{-i\omega t}$ we find, upon substitution into \eq{d15}, that $\omega$ must obey one the following dispersion relations
\ball{d17}
\omega= \omega_\pm = -\frac{i\sigma}{2}\pm \sqrt{k^2-\frac{\sigma^2}{4}}\,.
\gal
Thus, the general solution of the homogeneous equation \eq{d15}, which describes propagation of the initial conditions,  reads
\ball{d19}
a_{\b k}(t)= \alpha e^{-i\omega_+(t-t_0)}+\beta e^{-i\omega _-(t-t_0)}\,,
\gal
where $\alpha$ and $\beta$ are constants to be determined from the initial conditions \eq{d12} and \eq{d13}. 
 The particular solution due to the external current density is of the form $A_{2\b k}\propto \delta e^{-ik_z v t }$, where $\delta$ is found upon substitution into \eq{d10}:
\ball{d23}
\delta= \frac{ev}{k^2-k_z^2v^2-ik_zv \sigma}\,.
\gal
Thus, the general solution to \eq{d10} is 
\ball{d25}
\b A_{2\b k}= \unit z \left\{  \alpha e^{-i\omega_+(t-t_0)}+\beta e^{-i\omega _-(t-t_0)}+ \frac{ev}{k^2-k_z^2v^2-ik_zv \sigma}e^{-ik_z v t}\right\}\,.
\gal
Applying the initial conditions \eq{d12} and \eq{d13} we can fix $\alpha$ and $\beta$. The final result is 
\ball{d27}
\b A_{2\b k}= &\unit z \left\{ \delta \left[ \left(
\frac{\omega_--k_zv}{\omega_+-\omega_-}e^{-i\omega_+(t-t_0)}-\frac{\omega_+-k_zv}{\omega_+-\omega_-}e^{-i\omega_-(t-t_0)}\right)e^{-ik_zvt_0}+ e^{-ik_z v t}\right]\right.\nonumber\\
&
+ \frac{1}{i(\omega_+-\omega_-)}\Phi_{\b k}\left[-i\omega_-  e^{-i\omega_+(t-t_0)}+i\omega_+e^{-i\omega_-(t-t_0)}\right]\nonumber\\
& 
+\left.
\frac{1}{i(\omega_+-\omega_-)}\Psi_{\b k}\left[-  e^{-i\omega_+(t-t_0)}+e^{-i\omega_-(t-t_0)}\right]
\right\}\,.
\gal
Fourier transformation to the configuration space yields  exact analytical solution to the initial value problem \eq{d5}--\eq{d8}. Analytical and numerical evaluations of the integral over $\b k$ are challenging. Fortunately, in the ultra-relativistic limit  $\gamma\gg 1$, which is relevant for the relativistic heavy-ion collisions, expression for the vector potential \eq{d27} significantly simplifies \cite{Tuchin:2013apa}.
This is the subject of the next section.

\section{Diffusion approximation}\label{sec:b}

For an ultra-relativistic charge moving along the trajectory $z=vt$,  $\partial_t^2- \partial_z^2\sim k_z^2/\gamma^2\ll k_\bot^2, \sigma k_z$ implying that \eq{d5} can be approximated by 
\ball{b1}
\nabla_\bot^2 \b A_2(\b r, t)= \sigma \partial_t \b A_2(\b r, t)-\b j(\b r, t)\,.
\gal
This equation, being of the first order in time derivative, requires  only one initial condition
\ball{b2}
\b A_2(\b r, t_0)= \b A_1(\b r, t_0)= \unit z\Phi(\b r,t_0)\,.
\gal
We can solve the initial value problem \eq{b1}-\eq{b2} for an arbitrary time-dependence of the conductivity $\sigma(t)$. Introducing a new ``time"-variable $\lambda$ according to 
\ball{b4}
\lambda(t) = \int_{t_0}^t\frac{dt'}{\sigma(t')} 
\gal
and transferring \eq{b1} to the momentum space we obtain
\ball{b6}
-k_\bot^2\b A_{2\b k}= \partial_\lambda \b A_{2\b k}- \b j_{\b k}\,.
\gal
The corresponding  homogeneous equation (i.e.\ \eq{b6} with $\b j_{\b k}=0$) is solved by
\ball{b8}
\b a_{\b k}(\lambda)=\unit z Ce^{-k_\bot^2 \lambda}\,,
\gal
where $C$ is a constant. To derive a particular solution, we treat $C$ as a function of $\lambda$ and plug into \eq{b6}. We get
\ball{b10}
C= ev \int_0^\lambda d\lambda' e^{k_\bot^2\lambda'-ik_z v t(\lambda')}+D\,.
\gal
Substituting into \eq{b8} we find the general solution to \eq{b6}
\ball{b12}
\b A_{2\b k}(t)= \unit z \left\{ eve^{-k_\bot^2 \lambda}\int_0^\lambda d\lambda' e^{k_\bot^2\lambda'-ik_z v t(\lambda')}+De^{-k_\bot^2 \lambda}\right\}\,.
\gal
Since   $\lambda(t_0)=0$, the initial condition \eq{b2} implies that $D= \Phi_{\b k}(\b r, t_0)$. So finally,
\ball{b14}
\b A_{2\b k}(t)= \unit z \left\{ eve^{-k_\bot^2 \lambda(t)}\int_{t_0}^t \frac{dt'}{\sigma(t')} e^{k_\bot^2\lambda(t')-ik_z v t'}+\Phi_{\b k}e^{-k_\bot^2 \lambda(t')}\right\}\,.
\gal
In a particular case of constant electrical conductivity \eq{b14} simplifies to 
\ball{b16}
\b A_{2\b k}(t)= \unit z \left\{ \frac{ev}{\sigma}\frac{1}{\frac{k_\bot^2}{\sigma}-i k_z v}\left( e^{-ik_z v t}-e^{-\frac{k_\bot^2}{\sigma}(t-t_0)}e^{-ik_z v t_0}\right) + \Phi_{\b k}e^{-\frac{k_\bot^2}{\sigma}(t-t_0)}
\right\}\,.
\gal
This expression can be derived directly from \eq{d27}, but the approach described in this section is more straightforward. Fourier transformation to the configuration space 
\ball{b18}
\b A_2(\b r, t)= \int \frac{d^2k_\bot}{(2\pi)^2}\int_{-\infty}^{+\infty}\frac{dk_z}{2\pi}e^{i\b k_\bot\cdot \b b+i k_z z}\b A_{2\b k}(t)
\gal
can be done using the following integrals
\bal
&\int \frac{d^2k_\bot}{(2\pi)^2}\int_{-\infty}^{+\infty}\frac{dk_z}{2\pi}e^{i\b k_\bot\cdot \b b+i k_z z} e^{-k_\bot^2[\lambda(t)-\lambda(t')]}= \frac{\exp\left\{-\frac{b^2}{4[\lambda(t)-\lambda(t')]} \right\} }{4[\lambda(t)-\lambda(t')]}\delta(z-vt')\,, \label{b20}\\
&\int \frac{d^2k_\bot}{(2\pi)^2}\int_{-\infty}^{+\infty}\frac{dk_z}{2\pi}e^{i\b k_\bot\cdot \b b+i k_z z}
e^{-k_\bot^2\lambda(t)}\frac{ev}{k_z^2/\gamma^2+k_\bot^2}e^{-ik_z v t_0}\nonumber\\
&=\frac{\gamma e v}{4\pi}\int_0^\infty dk_\bot J_0(k_\bot b) e^{-k_\bot^2\lambda(t) - k_\bot \gamma|z-vt_0|}
\,.\label{b21}
\gal
Substituting \eq{b14} into \eq{b18}, doing integrals \eq{b20},\eq{b21} and then integrating over $t'$ yields
\ball{b23}
\b A_2(\b r, t)= &\frac{\unit z e}{4\sigma(z/v)}\frac{\exp\left\{-\frac{b^2}{4[\lambda(t)-\lambda(z/v)]} \right\} }{4[\lambda(t)-\lambda(z/v)]}\theta(tv-z)\theta(z-vt_0)\nonumber\\
&+ \frac{\gamma e v\unit z}{4\pi}\int_0^\infty dk_\bot J_0(k_\bot b) e^{-k_\bot^2\lambda(t) - k_\bot \gamma|z-vt_0|}\,.
\gal
Magnetic field can be calculated as in \eq{a23} with the following result
\ball{b26}
\b B_2= \b B_\text{val}+\b B_\text{init}\,,
\gal
where  the ``valence" $\b B_\text{val}$ and ``initial" $\b B_\text{init}$  components   are given by 
\bal
e\b B_\text{val}(\b r, t)=&\unit \phi \frac{\alpha \pi b }{2\sigma(z/v)[\lambda(t)-\lambda(z/v)]^2}\exp\left\{-\frac{b^2}{4[\lambda(t)-\lambda(z/v)]} \right\} \theta(tv-z)\theta(z-vt_0)\,,\label{b27}\\
e\b B_\text{init}(\b r, t)=&\unit \phi\gamma \alpha v\int_0^\infty dk_\bot k_\bot J_1(k_\bot b) \exp\left\{-k_\bot^2\lambda(t) - k_\bot \gamma|z-vt_0|\right\}\,. \label{b25}
\gal
The fine structure constant $\alpha = e^2/(4\pi)$.  Note that at $t=t_0$, $\b B_\text{val}$ vanishes whereas $\b B_\text{init}$ yields the initial condition \eq{a23}. $\b B_\text{init}$ is the field  that permits the plasma as it emerges at $t=t_0$ (at which time it coincides with $\b B_1$) and spreads in it according to \eq{b25}.  Unlike $\b B_\text{val}$, it strongly dependences on the collision energy $2\gamma$ (in units of proton mass). 

$\b B_\text{val}$ describes induced electromagnetic field generated as a response of QGP to electromagnetic field of the valence charge and builds up starting from $t=t_0$. 
Because of the two step-functions in \eq{b27} that reflect causality, $\b B_\text{val}$ is finite only in the interval $vt_0\le z\le vt$. In particular, it vanishes in the midrapidity $z=0$. At fixed $z$ satisfying $z\ge v t_0$,  $\b B_\text{val}$ emerges when $t= z/v$. An important property of  $\b B_\text{val}$  is that its magnitude is independent of energy (since $v\approx 1$).

At early times since the QGP creation viz.\ $t\gtrsim t_0$, expression in the exponent of \eq{b25} is such that  $k_\bot^2\lambda\ll k_\bot \gamma |z-vt_0|$ implying that  $\b B_\text{init}\approx \b B_1$. However, at later times when   $k_\bot^2\lambda\gg k_\bot \gamma |z-vt_0|$, we get 
\ball{b30}
e\b B_\text{init}= \unit\phi \frac{\gamma \alpha v b\sqrt{\pi}}{8\lambda^{3/2}}e^{-\frac{b^2}{8\lambda}}\left[ I_0\left( \frac{b^2}{8\lambda}\right) - I_1\left( \frac{b^2}{8\lambda}\right) \right]\,.
\gal
Since $k_\bot b\sim \sqrt{8}$ (which can be seen from $J_1$ series expansion) and $\lambda\sim (t-t_0)/\sigma$ we estimate that \eq{b30} is valid at times $t$ satisfying
\ball{b32}
\frac{t-t_0}{|z-vt_0|}\gg \frac{1}{\sqrt{8}}\gamma \sigma b\,.
\gal
At $z=0$, $b=7$~fm and $t_0=0.2$~fm this implies $t\gg 1$~fm, where I used $\sigma=5.8$~MeV known from the lattice calculations \cite{Ding:2010ga}, see also \cite{Aarts:2007wj,Amato:2013oja,Cassing:2013iz}.  Furthermore, since $b^2/8\lambda \ll 1$ we  expand \eq{b30} to obtain the late-time behavior of the initial magnetic field
\ball{b33}
e\b B_\text{init}\approx \frac{\gamma \alpha v\sqrt{\pi}b}{8\lambda^{3/2}}\unit \phi\,.
\gal
For constant $\sigma$ the late-time dependence (viz.\ $t\gg t_0$) is $ B_\text{init}\sim 1/t^{3/2}$. Notice that at late times  the ``valence" contribution  decays as $ B_\text{val}\sim 1/t^2$. It therefore emerges that the initial magnetic field dominates at early and late times.

\section{Magnetic field of two counter-propagating charges}\label{sec:c}

\begin{figure}[ht]
      \includegraphics[height=5cm]{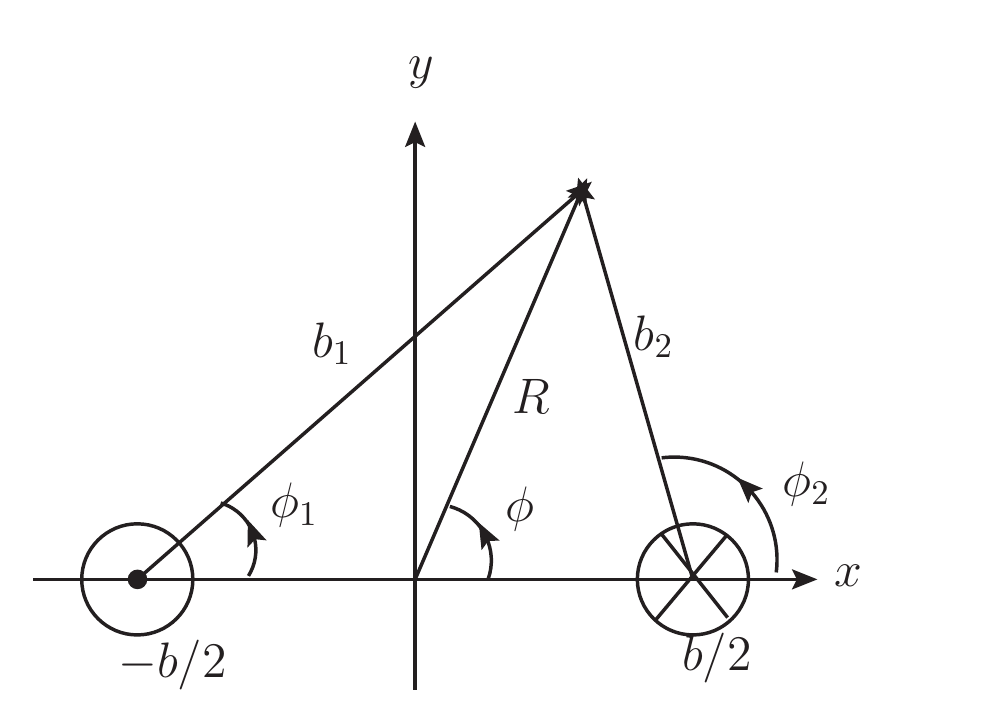} 
  \caption{Two counter-propagating charges $e$. One charges moves  along the positive $z$-axis at $z=vt$, $x=-b/2$, $y=0$ while another one moves in the opposite direction at $z=-vt$, $x=b/2$, $y=0$.}
\label{fig:c1}
\end{figure}

To calculate magnetic field in a heavy-ion collision one considers two sets of $Z$ counter-propagating  electric charges distributed according to one of the known nuclear density parameterizations, see e.g.\ \cite{Bzdak:2011yy}. However, to study the time evolution of magnetic field it suffices to consider just two counter-propagating charges.  The geometric symmetry of this configuration is similar to that of the event-average over many heavy-ion collisions at impact parameter $b$, but drastically reduces the computational time. The configuration that we consider is depicted in \fig{fig:c1}.

\begin{figure}[ht]
\begin{tabular}{cc}
      \includegraphics[height=5cm]{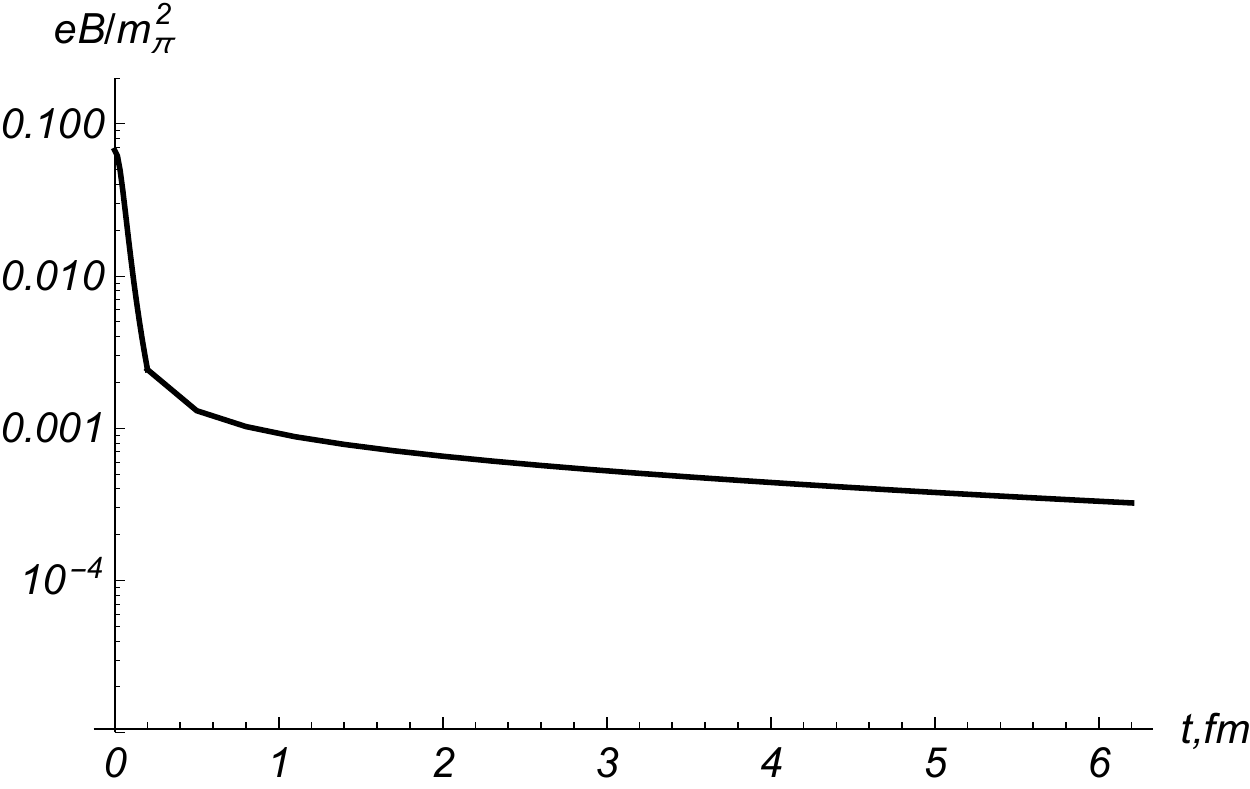} &
      \includegraphics[height=5cm]{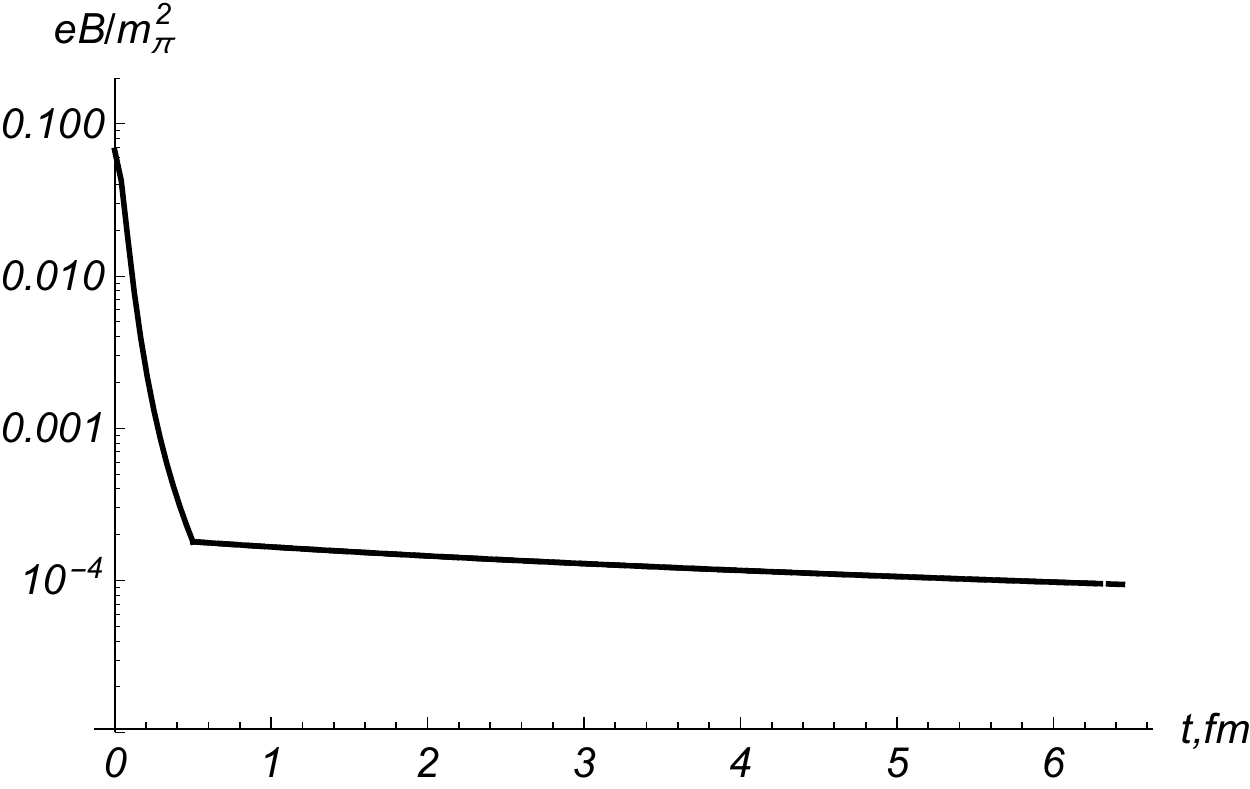}
            \end{tabular}
 \caption{Magnetic field in units of $m_\pi^2/e$. $\sigma=5.8$~MeV, $z=0$~fm ($\eta=0$). Left panel: $t_0=0.2$~fm, right panel:  $t_0=0.5$~fm.   Valence current does not contribute at all ($B_\text{val}=0$).  }
\label{fig:c2}
\end{figure}

\begin{figure}[ht]
\begin{tabular}{cc}
      \includegraphics[height=5cm]{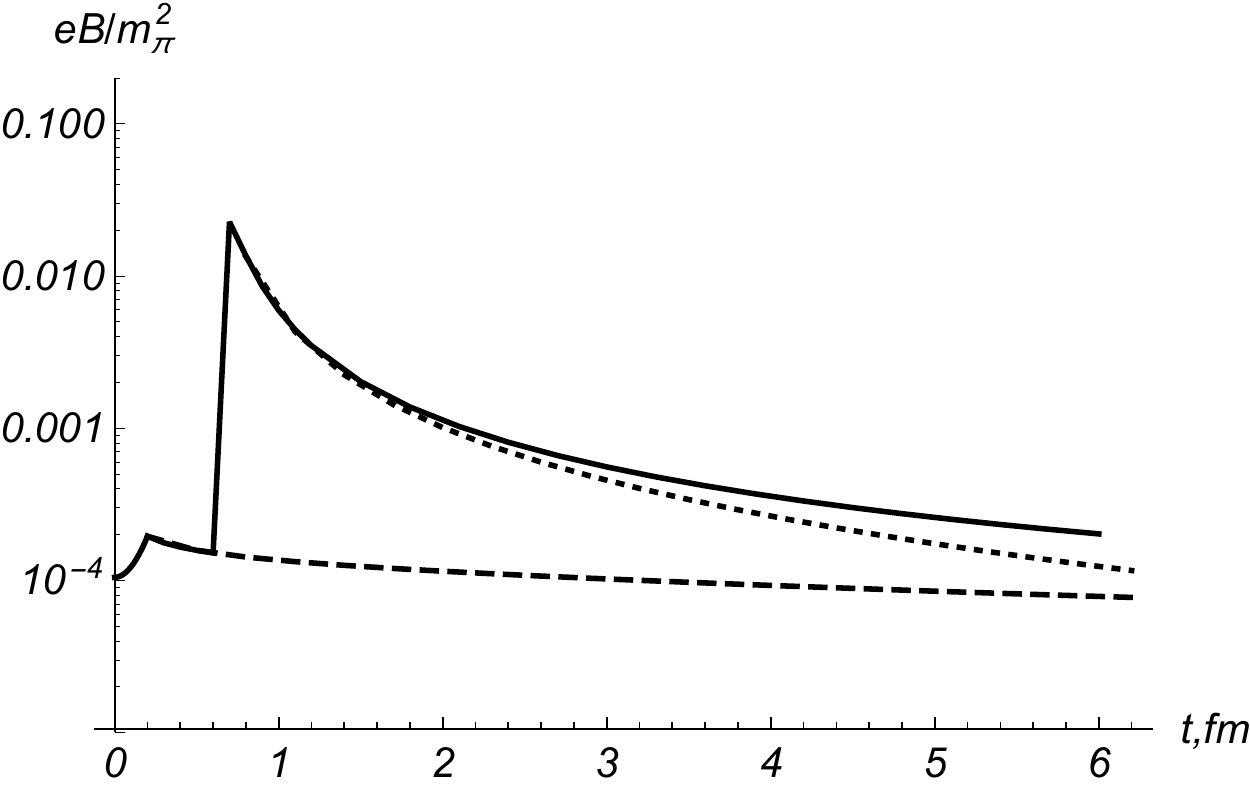} &
      \includegraphics[height=5cm]{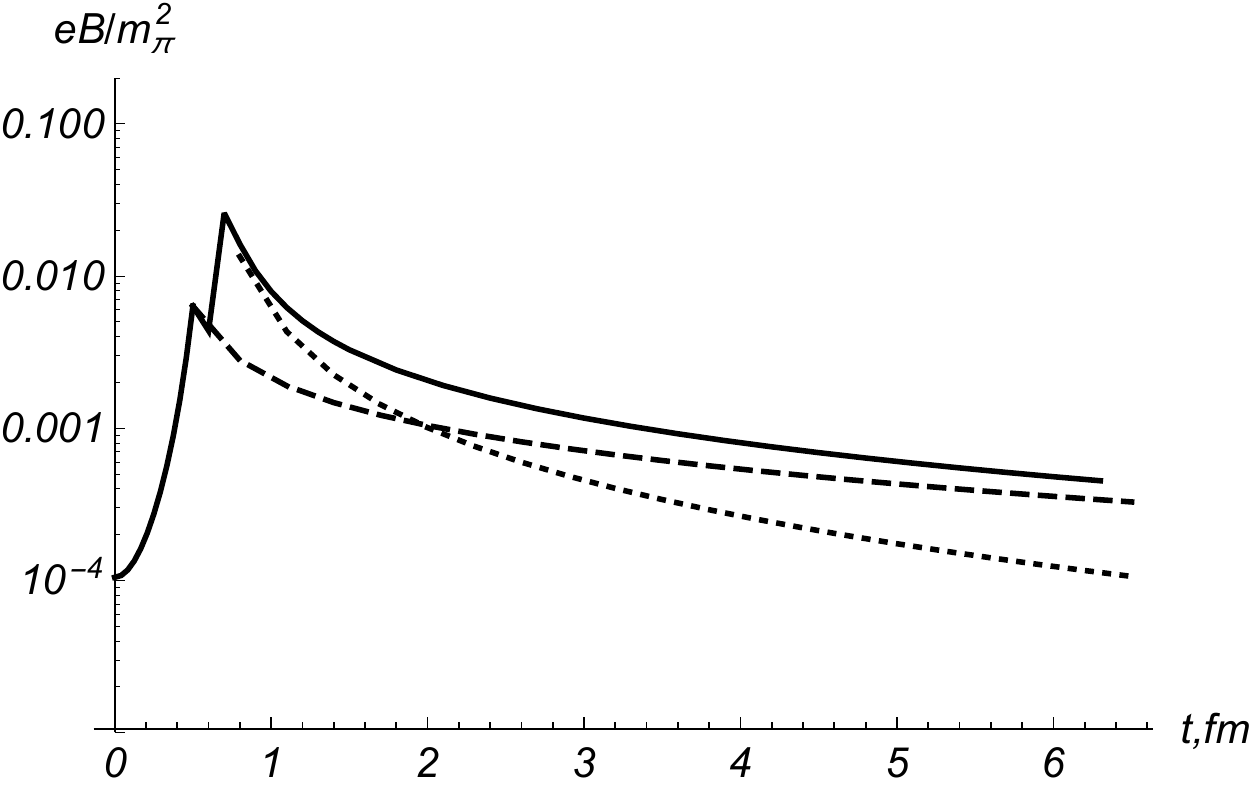}
            \end{tabular}
 \caption{Magnetic field in units of $m_\pi^2/e$.  $\sigma=5.8$~MeV, $z=0.6$~fm ($\eta=0.086$). Left panel: $t_0=0.2$~fm, right panel:  $t_0=0.5$~fm.  Solid, dashed and dotted lines stand for $B$, $ B_\text{init}$ and $ B_\text{val}$.}
\label{fig:c3}
\end{figure}

\begin{figure}[ht]
\begin{tabular}{cc}
      \includegraphics[height=5cm]{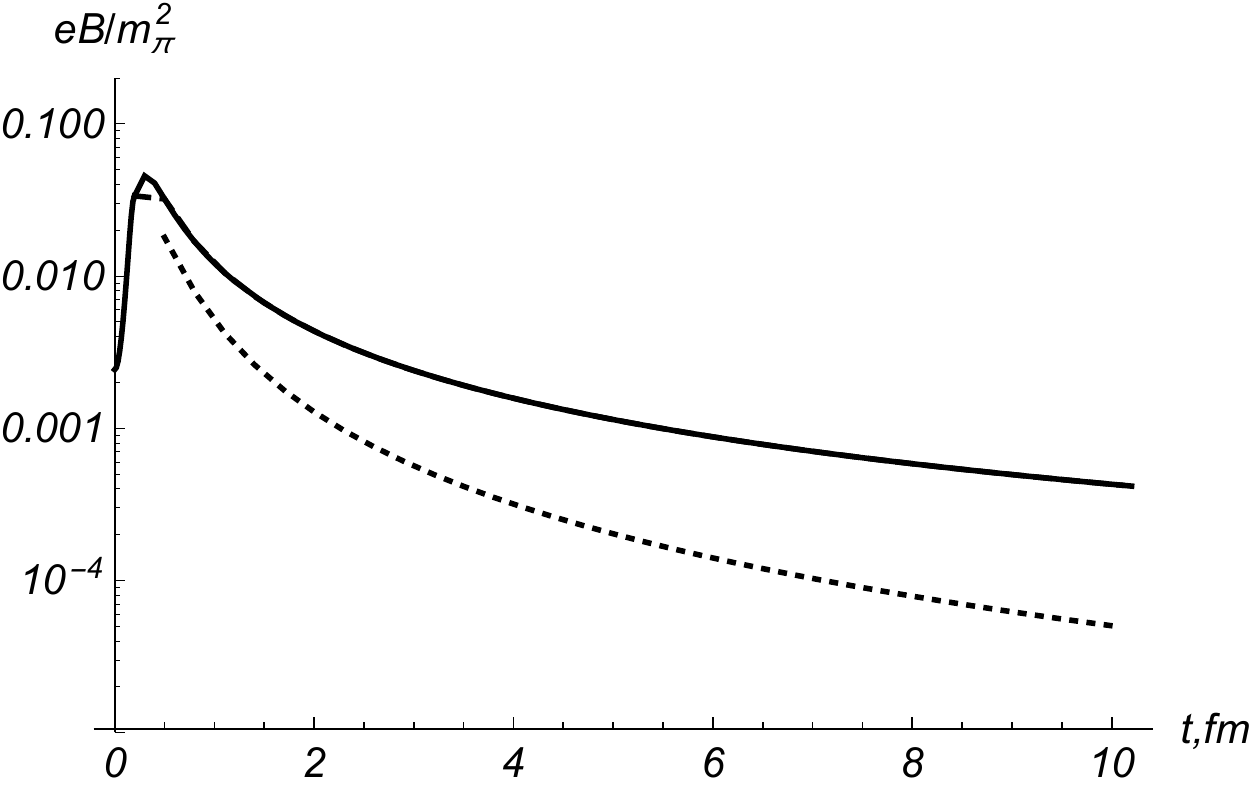} &
      \includegraphics[height=5cm]{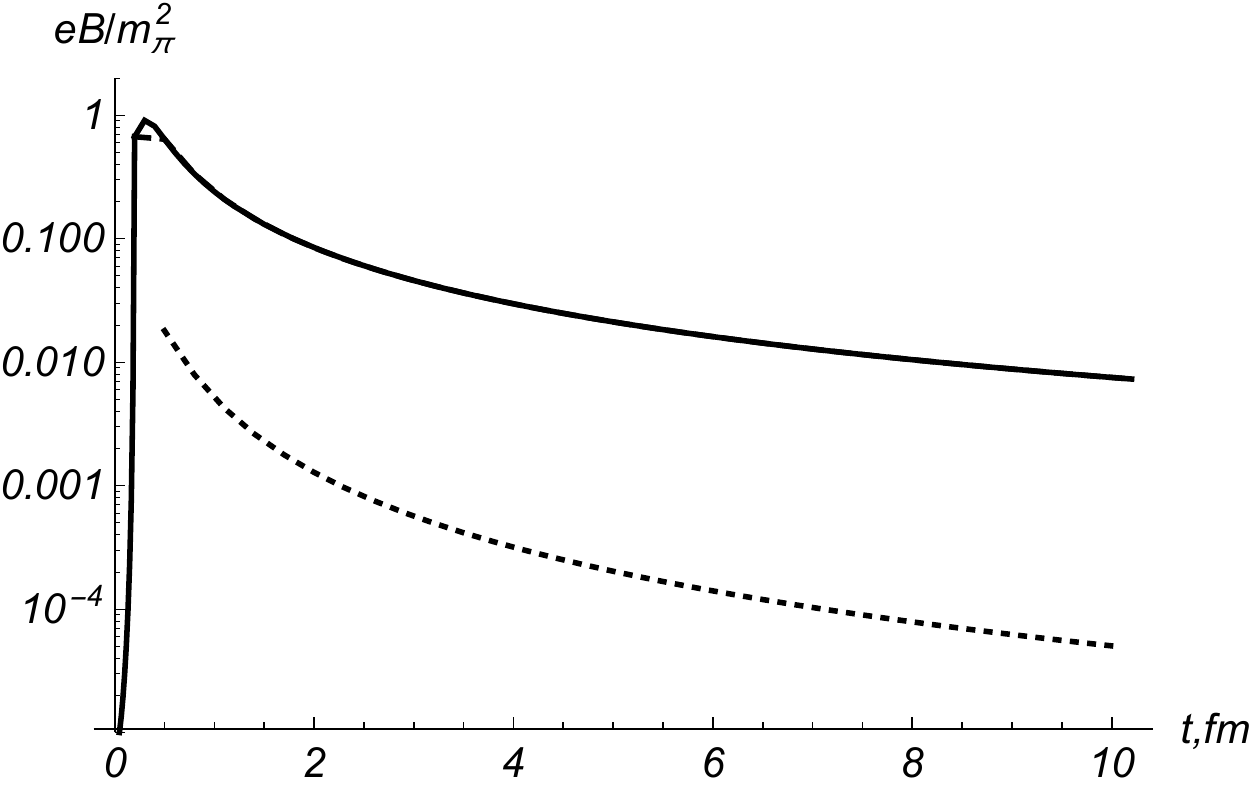} 
            \end{tabular}
  \caption{Magnetic field in units of $m_\pi^2/e$.   $\sigma=5.8$~MeV, $z=0.2$~fm $t_0=0.2$~fm. Solid, dashed and dotted lines stand for  $B$, $B_\text{init}$ and $B_\text{val}$. Left panel: $\gamma=100$ (RHIC), right panel: $\gamma=2000$ (LHC).}
\label{fig:c5}
\end{figure}

\begin{figure}[ht]
\begin{tabular}{cc}
      \includegraphics[height=5cm]{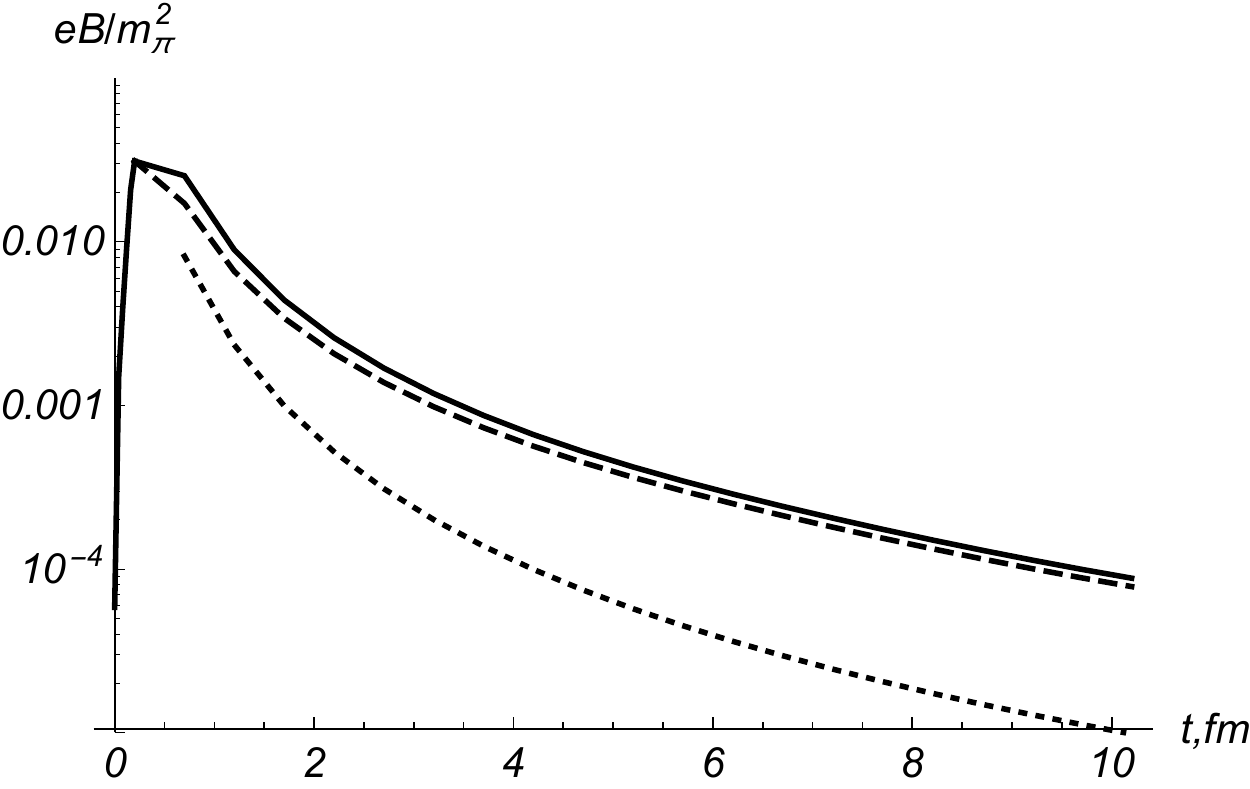} &
      \includegraphics[height=5cm]{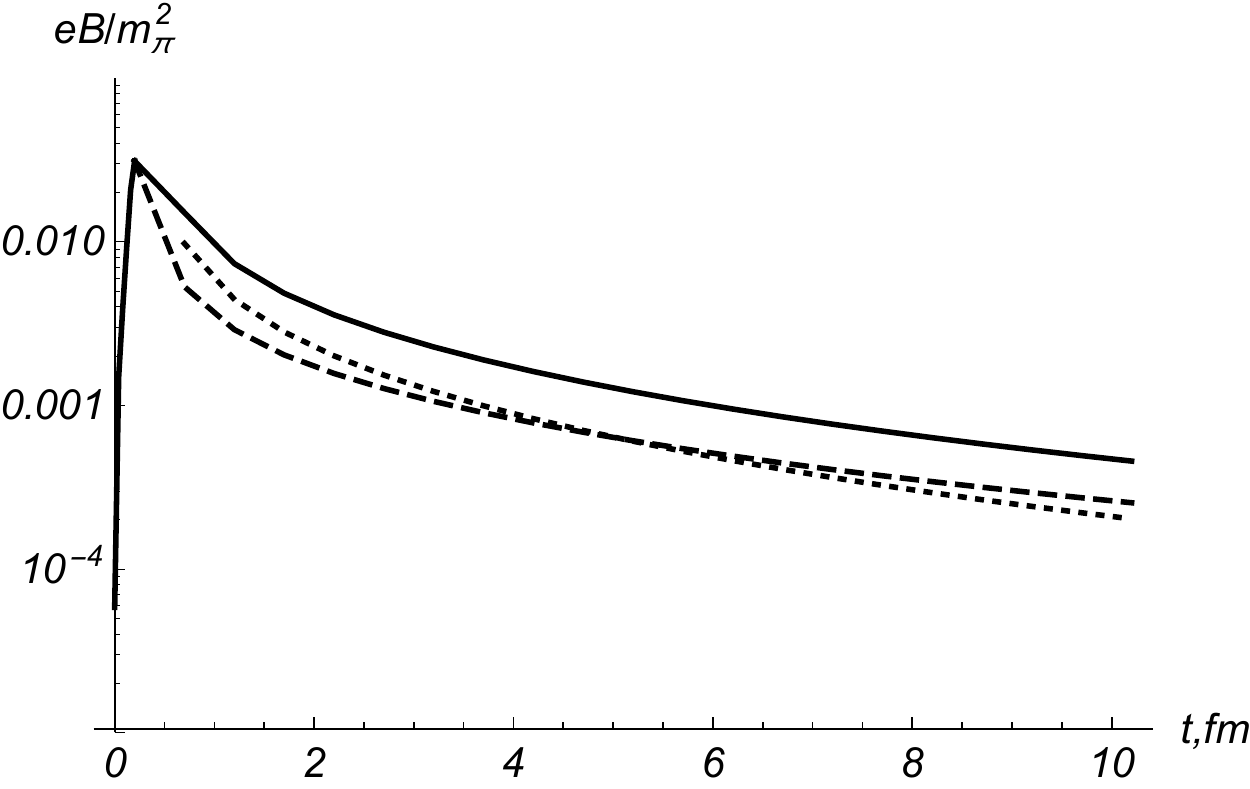}
            \end{tabular}
 \caption{Magnetic field in units of $m_\pi^2/e$. $z=0.2$~fm $t_0=0.2$~fm. Left panel: model A. Right panel: model B. Solid, dashed and dotted lines stand for $B$, $B_\text{init}$ and $B_\text{val}$.}
\label{fig:c4}
\end{figure}

Let $B^{(1)}(\b r_1,t)$ and $B^{(2)}(\b r_2,t)$ be magnitudes of the fields of the two charges, each given by  \eq{b26}--\eq{b25}. We can express coordinates of the observation point relative to each charge  $\b r_1=\b b_1+\unit z z_1$ and $\b r_2= \b b_2+\unit z z_2$ in terms of their center-of-mass in cylindrical coordinates $R,z,\phi$ as follows, see \fig{fig:c1}
\ball{c1} 
b_a= \sqrt{b^2/4+R^2+(-1)^a\, bR\cos\phi}\,,\quad \tan\phi_a= \frac{R\sin\phi}{R\cos\phi-(-1)^a\, b/2}\,,\quad z_a= vt+(-1)^a\, z\,.
\gal
where $a=1,2$ labels the charges. Noting that $\b B^{(a)}\propto \unit \phi_{a}$ and expressing $\unit \phi_{a}$ in terms of $\unit b$ and $\unit \phi$ we obtain  magnetic field in terms of the center-of-mass frame coordinates
\ball{c10}
\b B= &\unit b [ B^{(1)}(\b r_1,t)\sin(\phi-\phi_1)+B^{(2)}(\b r_2,t)\sin(\phi-\phi_2)]\nonumber\\
+&\unit \phi [ B^{(1)}(\b r_1,t)\cos(\phi-\phi_1)+B^{(2)}(\b r_2,t)\cos(\phi-\phi_2)]\,,
\gal
where $\b r_a$ and $\phi_a$ are replaced as indicated in \eq{c1}. 
The result is shown in  \fig{fig:c2}--\fig{fig:c4} in terms of a dimensionless and unit-independent quantity $eB/m_\pi^2$. In all figures impact parameter is $b=1$~fm, observation point is at  $\phi= \pi/2$, $R=7$~fm (i.e.\ $x=0$ and $y=7$ fm), and $\gamma=100$ (except \fig{fig:c5}). Also indicated is the pseudorapidity $\eta= -\ln[-(z/R)+\sqrt{(z/R)^2+1}]$.
Solid lines indicate the total magnetic field $B$, dashed lines represent the contribution of the initial condition $B_\text{init}$ and  dotted lines stand for the contribution of the valence charges $B_\text{val}$.  As discussed at the end of the previous section valence charge contribution decreases with time faster than that of the initial condition.

\fig{fig:c2}--\fig{fig:c5} depict magnetic field at constant electrical  conductivity $\sigma=5.8$~MeV \cite{Ding:2010ga}. In \fig{fig:c2} we compare magnetic field that is generated when QGP emerges at $t_0=0.2$~fm and at $t_0=0.5$~fm. Since magnetic field in vacuum decreases as $1/t^3$, see \eq{a23}, the late emergence of conducting medium means that the magnitude of the field in the former case is about 15 times larger than in the later. In both cases time-dependence of magnetic field in plasma is mild. Because of the step functions in \eq{b27} magnetic field at midrapidity $z=0$  is entirely due to the initial field $B_\text{init}$.

\fig{fig:c3} is similar to  \fig{fig:c2} except that $z=0.6$~fm unlocking the  ``valence" contribution. Being independent of the initial value of magnetic field at $t_0$ the  ``valence" contribution rapidly increases to its maximal value, that can be determined from \eq{b27} \cite{Tuchin:2013ie}. It then decreases at larger $t$ and becomes smaller than $B_\text{init}$.  Sharp lines seen in  \fig{fig:c3} indicate that the transition dynamics near $t=t_0$ is not fully captured by the diffusion approximation.

Energy dependence of magnetic field between the RHIC and LHC energies can be seen in  \fig{fig:c5}. $B_\text{init}$ grows approximately proportional to the collision energy $\gamma$, whereas $B_\text{val}$ is energy independent. Thus, at the LHC  magnetic field induced by valence charges is negligible. 

So far we considered only the case of constant electrical conductivity.  In practice, however electrical conductivity is time-dependent. To see the impact of $\sigma$ time-dependence on the time evolution of magnetic field we consider two models. In model A we assume that QGP emerges instantly at $t=t_0$ with $\sigma =5.8$~MeV and then cools down as it expands according to the Bjorken scenario \cite{Bjorken:1982qr}.  Namely, expansion is supposed to be  isentropic $nV= \text{const}$, where $n$ is the particle number density and $V$ is plasma volume. Since $n\sim T^3$ and at early times expansion is one-dimensional $V\sim t$ it follows that    $T\propto t^{-1/3}$.  Since $\sigma(t)\propto T$ we conclude that  $\sigma(t)\sim t^{-1/3}$. Thus a reasonable model for time dependence of electrical conductivity is 
\ball{c2}
 \sigma(t)= \frac{ \sigma }{2^{-1/3}(1+t/t_0)^{1/3}}\,, \quad \text{Model A}.
 \gal 
Another possibility is that the QGP does not appear as a thermal medium right away at $t=t_0$, rather it takes time $\tau$ until the conductivity reaches its equilibrium value $\sigma$. This can be described as 
\ball{c4}
\sigma(t)= \sigma\left(1-e^{-t/\tau}\right)\,,\quad  \text{Model B}.
\gal
We set conservatively $\tau =1$~fm. Note that we cannot let $\sigma(t)$ vanish at $t=t_0$ because that would violate the diffusion approximation that lead to \eq{b1}. However, \eq{c4} insures that $\sigma(t_0)\ll \sigma$. 

 In \fig{fig:c4} we contrast the two models. Similar calculation  at constant conductivity is shown in the left panel of  \fig{fig:c5}. We observe that time-dependence \eq{c2} (model A) significantly reduces magnetic field at later times. As far as model B is concerned,  time dependence \eq{c4} affects mostly $B_\text{val}$ because it directly depends on $\sigma(t)$, whereas $B_\text{init}$  depends only on $\lambda(t)$, see \eq{b27},\eq{b25}. Model B has minor effect  on the total magnetic field, although one can certainly find regions in space-time where its effect is more pronounced.  What actually matters is the initial time $t_0$ at which one can treat the produced particle system as a medium. As long as conductivity is large enough at later times, magnetic field is fairly insensitive to the precise QGP dynamics.  

\section{Summary}\label{sec:f}

Just before the QGP emerges, the interaction region is permitted by the primordial electromagnetic field created by valence charges of two heavy-ions. At the initial time $t_0$ this magnetic field smoothly connects to the magnetic field in plasma and evolves according to the Maxwell equations in the electrically conducting medium. In addition to this ``initial" magnetic field, there is another ``valence" contribution that arises from the external valence electric charges inducing currents in the QGP. It has been tacitly assumed  that the former contribution is not important  \cite{Tuchin:2013apa}. In this paper we argued to the contrary, that the initial magnetic field dominates at very early and later times and increases much faster with the collision energy than the ``valence" contribution. 

We also studied the effect of time dependence of  electrical conductivity and concluded that at early times it has a rather minor effect on the field strength, as long as the produced particle system can be treated as a medium at early enough time. However, towards the later times of plasma evolution, time-dependence of electrical conductivity plays an important role. In the Bjorken scenario it leads to much weaker fields as compared to the constant conductivity case. 

We considered the case of two counter-propagating charges that gives an accurate picture for the time dependence of the event-averaged fields in heavy-ion collisions. Scaling the result with $Z$ we can obtain an estimate of the  magnetic field strength in heavy-ion collisions. Calculating the spatial distribution requires an accurate account of  the exact nuclear geometry, which is not difficult using the results reported in this paper.

\acknowledgments
This work  was supported in part by the U.S. Department of Energy under Grant No.\ DE-FG02-87ER40371.



\end{document}